\documentclass[12pt]{article}
\usepackage{amsthm}
\usepackage{amsmath}
\usepackage{moreverb}
\usepackage{multirow}
\usepackage[colorlinks,bookmarksopen,bookmarksnumbered,citecolor=red,urlcolor=red]{hyperref}
\usepackage{subfigure}
\usepackage{rotating}
\usepackage{wrapfig}
\usepackage{natbib}
\usepackage{atbegshi,picture,afterpage}
\usepackage{placeins} 
\newcommand{\iid}  {\stackrel{\rm iid}{\sim}}
\newcommand\BibTeX{{\rmfamily B\kern-.05em \textsc{i\kern-.025em b}\kern-.08em
T\kern-.1667em\lower.7ex\hbox{E}\kern-.125emX}}

\begin{document}


\title{A Bayesian Joint model for Longitudinal DAS28 Scores and Competing Risk Informative Drop Out in a Rheumatoid Arthritis Clinical Trial}

\author{Violeta Hennessey
\thanks{Warner Home Video,  Los Angeles, CA, USA. E-mail: v.g.hennessey@gmail.com}\and
Luis Le\'on Novelo\thanks{University of Texas Health Science Center at Houston- School of Public Health, Houston, TX, USA}
\and 
Juan Li \thanks{Department of Biostatistics, Amgen Inc., Thousand Oaks, CA, USA}
\and Li Zhu \footnotemark[3]\and Xin Huang \footnotemark[3]
\and
 Eric Chi\thanks{Department of Biostatistics, Amgen Inc.,  South San Francisco, CA, USA}
 \and
 Joseph G. Ibrahim \thanks{Department of Biostatistics, The University of Noth Carolina at Chapel Hill, Chapel Hill, NC, USA}}


\maketitle
\begin{abstract}
Rheumatoid arthritis clinical trials are strategically designed to collect the disease activity score of each patient over multiple clinical visits, meanwhile a patient may drop out before their intended completion due to various reasons. The dropout terminates the longitudinal data collection on the patients activity score.   In the presence of informative dropout, that is, the dropout depends on latent variables from the longitudinal process, simply applying a model to analyze the longitudinal outcomes may lead to biased results because the assumption of random dropout is violated. In this paper we develop a data driven Bayesian joint model for modeling DAS28 scores and competing risk informative drop out. The motivating example is a clinical trial of Etanercept and Methotrexate with radiographic Patient Outcomes  \citep[TEMPO][]{keystone2009patients}.
\end{abstract}


\afterpage{\AtBeginShipout{\AtBeginShipoutUpperLeft{%
  \put(\dimexpr\paperwidth-27mm\relax,-1.5cm){\makebox[0pt][r]{\textit{Hennessey et.al.}}}%
}}}


\section{Introduction}
Rheumatoid arthritis is a chronic inflammatory disease that affects the joints. Disease activity leads to pain, swelling, and joint damage. The cause of rheumatoid arthritis is unknown but it is known to be an autoimmune disease. Disease activity score based on 28 joints (DAS28) is a composite score measuring disease activity that integrates measures of physical examination (number of tender and swollen joints), erythrocyte sedimentation rate (a hematology test that is a non-specific measure of inflammation), and patient reported outcome (patient self reported global health on a visual analogue scale).
$$
\begin{array}{rl}
DAS28=&0.56\sqrt{\textrm{Tender28}}+0.28\sqrt{\textrm{Swollen28}}+0.70\log(\textrm{ESR})\\
&+0.014(\textrm{Patient Global Health on VAS})
\end{array}
$$
DAS28 scores can range from 0.49 to 9.07. A score greater than 5.1 is considered high disease activity, 3.2-5.1 moderate disease activity, 2.6-3.2 low disease activity, and a score less than 2.6 is considered to be clinical remission. The goal of treatment is to achieve the lowest possible level of disease activity and clinical remission if possible. Figure 1 displays the 28 joints considered in the calculation of DAS28 scores.

\begin{wrapfigure}{r}{0.4\textwidth}
 \begin{center}
 \includegraphics[width=.3\textwidth]{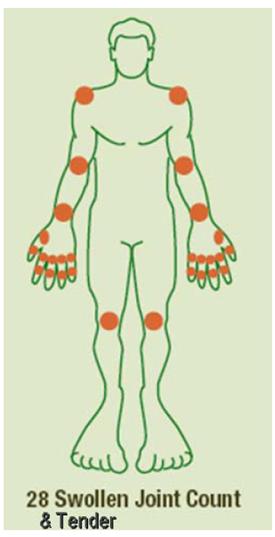}
 \end{center}
  \label{fig:fig2}
\label{fig:subfigureExample} \caption[]{Twenty-eight joints considered in the calculation of DAS28 scores.}
\end{wrapfigure}

Clinical trials are strategically designed to collect the disease activity score of each patient over multiple clinical visits, meanwhile a patient may drop out before their intended completion due to various reasons. The dropout terminates the longitudinal data collection on the patient's disease activity score. In the presence of informative dropout 
\citep{rubin1976inference,little1987statistical}, 
 that is, the dropout depends on latent variables from the longitudinal process, simply applying mixed effects models 
 to analyze longitudinal outcomes may lead to biased results because the assumption of random dropout is violated
 \citep{laird1982random}.  
 It would be more appropriate to jointly model the DAS28 scores and the dropout to account for the dependency between the two processes.

Several methods have been proposed in the literature for jointly modeling longitudinal and time-to-event data, as reviewed by 
\cite{hogan1997mixture,
tsiatis2004joint, 
yu2004joint, 
ibrahim2005bayesian,
hanson2011predictive}. 
 Joint models have been widely used in HIV/AIDS clinical trials, cancer vaccine (immunotherepy) trials and quality of life studies. References include 
\cite{
schluchter1992methods,
pawitan1993modeling,
de1994modelling,
tsiatis1995modeling,
faucett1996simultaneously,
wulfsohn1997joint,
faucett1998analysis,
henderson2000joint,
wang2001jointly,
xu2001evaluation, 
law2002joint,
song2002semiparametric,
chen2002bayesian,
chen2004new,
r2003bayesian,
brown2003bayesian, 
brown2005flexible,  
 chi2006joint,chi2007new,                      
 zhu2012bayesian}, 
 and many others.

Our approach combines different methodologies to deal with the complexity of the data.
The joint model has two parts: the first is a longitudinal random change point model for repeated measures DAS28 scores; the second part is a time to event hazard risk log normal model that models two competing risk events. Both parts are linked through a patient specific random effect. Similar approaches are described next.
\cite{jacqmin2006random} propose a joint model for repeated measurements (cognitive test score) and time to event (dementia). The part of their model for repeated measures, as ours, assumes a linear trend before a random change point and, in contrast to ours, assumes a polynomial trend after it. They assume a lognormal model for the time to event as we do for our competing risk model.
\cite{faucett2002survival} apply a similar random effects model for the AIDS trial along with a time-dependent proportional hazards model for the time to AIDS. In their random effects longitudinal model they 
assume, as we do,
two linear trends before and after a random change point. 
Although, as ours, their model contains a subject-specific random intercept, in contrast to ours, all subjects have a common transition point. 
\cite{pauler2002predicting} adopts a similar random effect for the longitudinal PSA trajectories and Cox proportional hazard model for the progression times. By comparing the data pattern of people who relapse and who do not, they add a strong constraint to the random effects model with a new indicator – the slope after the transition point has to be greater than the slope before the transition point. 
For event-time data, Cox proportional hazards model with a piecewise exponential model is applied to increase flexibility.

Currently, the most commonly used method for jointly modeling lontidudinal and time-to-event data are likelihood based joint models. Likelihood based joint models are classified as either selection models or pattern mixture models \citep{little1995modeling}. 
Selection models factor the joint distribution through the marginal distribution model of the longitudinal outcome and the conditional distribution of time-to-event given longitudinal latent variables. Pattern-mixture models 
\citep{glynn1986selection,little1993pattern} 
factor the joint distribution differently by modeling the longitudinal data stratified by time-to-event (e.g., time-to-dropout) patterns. Our interest in this paper is towards marginal inference on the longitudinal profiles. 
This is, we are interested in 
assessing the treatment effects while adjusting for the imbalanced 
drop out.
 For this reason we consider selection models over pattern-mixture models. 
 We also consider a competing risk survival model for informative dropout because, in practice, a patient may drop out before their intended completion date due to various competing risks. Joint analysis of longitudinal measurements and competing risks failure time data has been studied recently both in terms of frequentist approaches 
 \citep{elashoff2007approach,elashoff2008joint} 
 and Bayesian approaches 
 \citep{
 huang2010joint}. 
 Such approaches of joint modeling can make the estimates for the longitudinal marginal profile less biased. Therefore we can draw more reliable conclusions from clinical trials with longitudinal outcomes.

The motivation comes from a randomized double-blinded clinical Trial of Etanercept and Methotrexate with radiographic Patient Outcomes  \citep[TEMPO, ][]{keystone2009patients}. 
The primary objective of this trial is to evaluate the efficacy and safety of a combination therapy of methotrexate and etanercept for the treatment of rhuematoid arthritis versus methotrexate alone and etanercept alone. Methotrexate, a disease-modifying antirheumatic drug (DMARD), indirectly inhibits the enzyme adenosine deaminase, resulting in anti-inflammatory effects. Etanercept, a tumor necrosis factor (TNF) inhibitor, is a large molecule biologic therapy that acts like a sponge to remove TNF-$\alpha$ molecules from the joints and blood, thereby reducing disease activity and slowing joint destruction. The two drugs have different mechanisms of action, therefore it is hypothesized that the combination therapy will be more effective in the treatment of rheumatoid arthritis. In the TEMPO study, a total of 682 subjects were randomized at a 1:1:1 ratio of methotrexate alone, etanercept alone or the combo-therapy. At the end of the three year study, approximately 52\% of the subjects dropped out from the study due to the following reasons: inefficacious treatment, adverse events, administrative decisions, and others (Table \ref{tab:summarystats}). Dropout due to inefficacious treatment was highest in the methotrexate group (18.4\%) compared to etanercept (17.5\%) and the combination treatment (5.6\%). Dropout due to adverse events was lowest in the combination treatment (20\%) compared to methotrexate (25.4\%) and etanercept (20.2\%).
\begin{table}
\caption{\label{tab:summarystats}Summary of subject disposition in TEMPO study}
\begin{center}
\begin{tabular}{lcccc}
& Methotrexate &Etanercept& Combo-&Total \\
&  alone           &        alone& therapy&     \\
& N=228 & N=223&N=231&N=682\\
\hline
Completed study& 88 (38.6\%)&107 (48.0\%)&132 (57.1\%)&327 (47.9\%)\\
\hline
{Discontinued study}& 140 (61.4\%)& 116 (52.0\%)&99 (42.9\%)&355 (52.1\%)\\
due to \\
\hspace{3mm}adverse event& 58 (25.4\%)&45 (20.2\%)&46 (20\%)&149 (21.8\%)\\
\hspace{3mm}inefficacious treatment& 42 (18.4\%)&39 (17.5\%)&13 (5.6\%)&94 (13.8\%)\\
\hspace{3mm}administrative decision& 9 (4.0\%)&13 (5.8\%)& 19 (8.2\%)&41 (6\%)\\
\hspace{3mm}other reasons& 31 (13.6\%)&19 (8.5\%)&21 (9.1\%)&71 (10.4\%)\\
\end{tabular}
\end{center}
\end{table}

The remainder of this paper is organized as follows. Section 2 presents our proposed Bayesian joint model. In Section 3, we apply the model to the TEMPO study and concluding remarks are provided in Section 4.
\section{Bayesian Joint Model}

\subsection{Model for Longitudinal DAS28 Scores}
Let $\mathbf{y}_{i}=(y_{i1},\ldots,y_{in_i})$ denote the DAS28 scores for subject $i$ where $n_i$ is the number of DAS28 scores collected over time for subject $i$ in the study. Let $\mathbf{t}_{i}=(t_{i1},\ldots,t_{in_i})$ denote the corresponding vector of observation times and $\mathbf{x}_{i}=(x_{i1},\ldots,x_{im})$ denote the covariate vector for subject $i$.

In the subject level data, we observed a change-point pattern in the DAS28 scores with a steep decrease in DAS28 scores after initial treatment followed by stability or a more gradual change. For this reasons we consider the following random change-point model
$$\log(y_{ij})=\psi_{ij}+\epsilon_{ij}$$

\begin{equation}\label{eq:logoflongitudinal}
\psi_{ij}=\alpha_{i}+ \beta_{1i}\kappa_i+\beta_{1i}(t_{ij}-\kappa_i)_- +\beta_{2i}(t_{ij}-\kappa_i)_+
\end{equation}
where $\psi_{ij}=\psi_i(t_{ij})$ is the trajectory function on the log scale for subject $i$ and $\epsilon_{ij}$ is the associated random residual error. We modeled $y_{ij}$ on the log scale to improve linearity and to remove the constraint of positive values. The trajectory function on the log scale corresponds to two straight lines that meet at a random change-point $\kappa_i$.
The parameters $\alpha_i$, $\beta_{1i}$, $\beta_{2i}$, and $\kappa_i$ represent the subject level intercept, slopes before and after the random change-point respectively. Here we use the notation $(t_{ij}-\kappa_i)_- = \min\{t_{ij}-\kappa_i, 0\}$ and
$(t_{ij}-\kappa_i)_+ = \max\{t_{ij}-\kappa_i, 0\}$. The random residual errors are assumed to be 
independent and
normally distributed with mean zero and variance $\sigma^2$.

We use normal prior distributions for the parameters $\alpha_i$, $\beta_{1i}$, $\beta_{2i}$, and $\kappa_i$.

$$\begin{array}{r l rl}
\alpha_{i}\sim& N(\mu_{\alpha_{i}}, \sigma^2_{\alpha}), &\mu_{\alpha_{i}}=&\gamma_{\alpha0}+\gamma_{\alpha1}x_{i1}+\ldots+\gamma_{\alpha m}{x}_{im}\\
\beta_{1i}\sim& N(\mu_{\beta_{1i}}, \sigma^2_{\beta_1}),  &\mu_{\beta_{1i}}=&\gamma_{\beta10}+\gamma_{\beta11}x_{i1}+\ldots+\gamma_{\beta1m}{x}_{im}\\
\beta_{2i}\sim& N(\mu_{\beta_{2i}}, \sigma^2_{\beta_2}),  &\mu_{\beta_{2i}}=&\gamma_{\beta20}+\gamma_{\beta21}x_{i1}+\ldots+\gamma_{\beta2m}{x}_{im}\\
\kappa_{i}\sim& N(\mu_{\kappa_{i}}, \sigma^2_{\kappa}), &\mu_{\kappa_{i}}=&\gamma_{\kappa0}+\gamma_{\kappa1}x_{i1}+\ldots+\gamma_{\kappa m}{x}_{im}\\
\end{array}
$$
We let the means of the parameters vary for each treatment group by letting $x_{i1}, \ldots, x_{im}$ be dummy variables for treatment. The $\gamma$ parameters are then coefficients that will represent treatment effect. 
In our application, with three treatment groups, $m=2$.
We complete the model by placing normal hyper-priors with means of zero and large variances on the hyperparameters $\gamma$, this is $N(0,10^3)$. 
 We also choose vague priors for the variances:
$\sigma^2,
\sigma^{2}_\alpha,\sigma^{2}_{\beta_1},\sigma^{2}_{\beta_2},
\sigma^{2}_{\kappa}\iid \text{Inverse gamma}(0.01,0.01)$.
We note that the above model is equivalent to a mixed-effect model where the "random-effect parameters" are re-parameterized and centered around the "fixed-effect parameters". This hierarchical centering reparameterization technique reduces autocorrelation between consecutive 
Markov chain Monte Carlo~(MCMC) samples 
\citep{gelfand1995efficient}.

\subsection{Model for Competing Risk Dropout}
Competing risk is defined by 
\cite{gooley1999estimation} 
as the situation where one type of event precludes the occurrence or alters the probability of occurrence of another event 
\citep{pintilie2006competing}. 
We consider the following competing risk model for informative dropout. For the motiviating example, dropout due to adverse event and dropout due to inefficacious treatment is considered to be informative. Dropout due to administrative decisions and other are considered uninformative, that is, dropout is random and does not depend on disease activity outcome.

Let $d_{ik}$ denote the time-to-dropout for subject $i$ due to risk $k$ where $k=1$ for adverse event and $k=2$ for inefficacious treatment. Let $c_{ik}$ denote the corresponding censoring indicators for subject $i$ where $c_{ik}=1$ if subject $i$ dropped out of the study due to risk $k$, 0 otherwise. Because dropout tends to occur in the earlier part of the study, we use a lognormal survival regression model for $d_{ik}$ to allow for a humped-shaped hazard function,
$$
\begin{array}{c}
\log(d_{ik})\sim \textit{N}(\theta_{ik},\varsigma^2_k)\\
\theta_{ik}=\varphi_{k0}+\varphi_{k1}x_{i1}+\ldots+\varphi_{km}{x}_{im} + \omega_{k}\nu_{i}
\end{array}
$$
\noindent Here $\varphi_{k0},\dots,\varphi_{km}$ are the cause-specific effects on dropout, $x_{i1}, \ldots, x_{im}$ are the dummy variables for treatment. The last term, $\omega_{k}\nu_{i}$, is the component for the joint analysis where $\nu_i$ is the latent variable (or vector)
that,
in order to link the longitudinal and dropout processes,
  is a function of the parameters in the RHS of 
\eqref{eq:logoflongitudinal}, and
$\omega_{k}$ is the coefficient (or vector of coefficients). See next section for more
detail about $\nu_i$.
  We use normal hyper-priors for $\varphi_{kj}$ and $\omega_{k}$ with means of zero and large variances, say 1000. 
We also assume a vague prior for $\varsigma^2_k\sim \text{Inverse gamma}(.01,.01)$.

\section{Application to the TEMPO Study}
We coded the joint model described in Section 2 in WinBUGS 
\citep[][see Appendix for code]{lunn2000winbugs}. 
 The covariates in the model are $x_{i1}$ and $x_{i2}$ representing the two dummy variables required for three treatment groups. For example, $x_{i1}=1$ if subject $i$ received methotrexate alone, 0 otherwise, and $x_{i2}=1$ if subject $i$ received etanercept alone, 0 otherwise. Here, the combo-therapy is the reference group.

For $\nu_{i}$, we consider different latent variables described in the literature. In the works of 
\cite{faucett1996simultaneously,
wang2001jointly,
ibrahim2004bayesian}, 
and 
\cite{r2003bayesian},  missingness (or dropout) is considered to be associated with outcome. This is known as an outcome dependent selection model where the estimated trajectory at the time of dropout is used as the latent variable. 
\cite{
wu1988estimation,
wulfsohn1997joint, 
follmann1995approximate,
henderson2000joint,
guo2004separate} 
considered an alternative approach known as the shared parameter selection model where the missingness (or dropout) is directly related to a trend over time. In the shared parameter selection models, the parameters from the trajectory function are used as the latent variables.  We considered the different latent variable(s) in the literature and used a selection model criterion, Deviance Information Criterion (DIC), to determine which latent variable(s) to use in the final model. The DIC provides a measure of goodness of fit penalized by the effective number of parameters. The model with the smallest DIC is preferred \citep{spiegelhalter2002bayesian}. 

For each model we ran 10,000 MCMC iterations with two chains, each with different initial values, retaining every fifth sample. We discarded the first 5,000 iterations as burn-in. We assessed convergence using standard diagnostic tools, that is, converging and mixing of the two chains. We also assessed autocorrelation as a function of time-lag.  Model 1 sets $\nu_{i}=0$, representing separate analysis where the longitudinal model and the dropout process are not linked. Model 1 assumes that dropout is not informative of outcome. Model 2 considers dropout is associated with the outcome at the time of dropout, denoted as $\psi_i^*$, $\nu_{i}=\psi_i^*$. Model 3 considers dropout is associated with the disease activity at baseline, $\nu_{i}=\alpha_i$. In Model 4, dropout is associated with the initial change in disease activity prior to the change-point, $\nu_{i}=\beta_{1i}$. In Model 5, dropout is associated with the change in disease activity following the change-point, $\nu_{i}=\beta_{2i}$. Model 6 shares all parameters from the trajectory function where $\nu_{i}$ is now a vector with $\mathbf{\nu_{i}}=(\alpha_i, \beta_{1i}, \beta_{2i})$. Convergence and acceptable autocorrelation was met by all six models. Model 6 is preferred with the smallest DIC (Table \ref{table:DICcomparison}).

\begin{table}
\caption{Deviance Information Criterion (DIC) for models fitted to the TEMPO study data. Model 6 is preferred with the smallest DIC.}\label{table:DICcomparison}
\centering
\begin{tabular}{lcr}
Model: Type of Analysis & $\nu_i$ & DIC \\
\hline
\ 1: Separate analysis  &  0  & 781.43\\
Joint analysis where dropout is related to:\\
\ 2:  disease activity at time of dropout& $\psi^*_{i}$ &488.98 \\
\ 3: baseline disease activity & $\alpha_{i}$ &790.63 \\
\ 4: initial change in disease activity after treatment & $\beta_{1i}$  &574.10 \\
\ 5: the long-term change in disease activity&$\beta_{2i}$  &406.45 \\
Joint analysis,\\
\ 6:  sharing all parameters from the trajectory& $\alpha_{i}, \beta_{1i}, \beta_{2i}$ & 391.04\\
\end{tabular}
\end{table}

\begin{figure}[h] 
\begin{center} 
\begin{tabular}{rl}
\includegraphics[scale=.35]{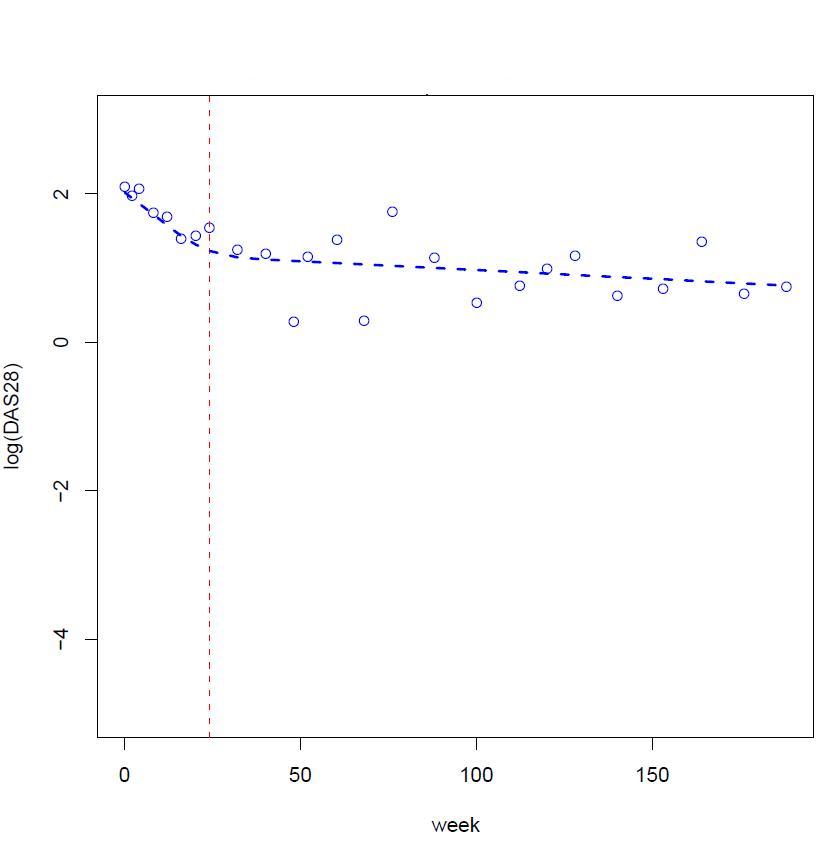}&\includegraphics[scale=.35]{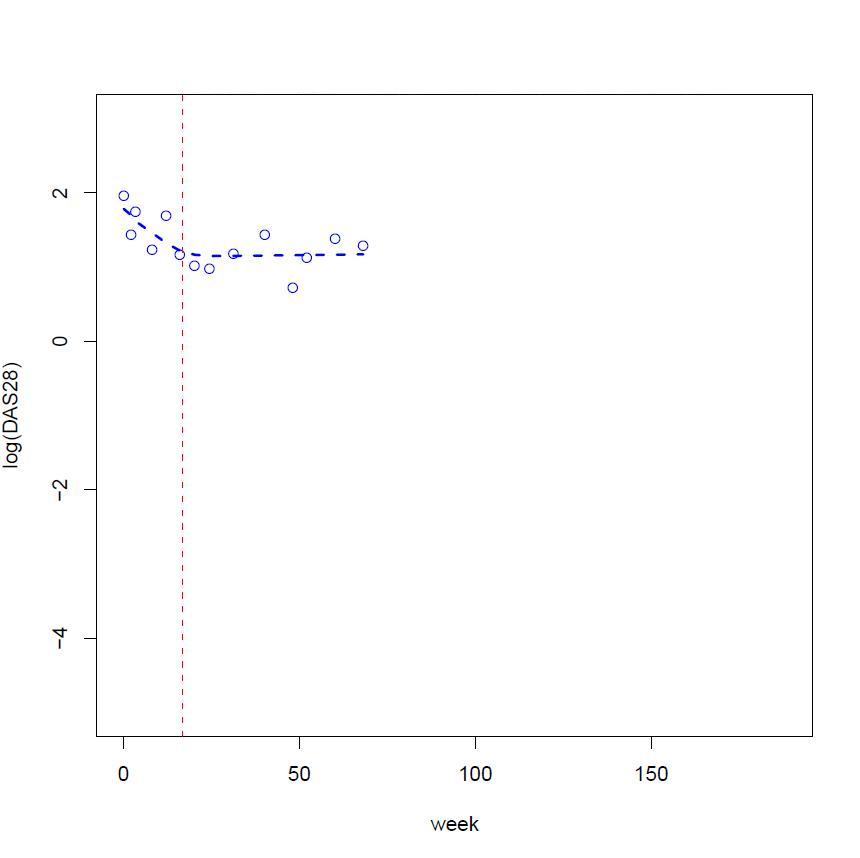}\\
\multicolumn{1}{l}{(a) Patient completed study} & (b) Patient dropped out due to AE\\
\includegraphics[scale=.35]{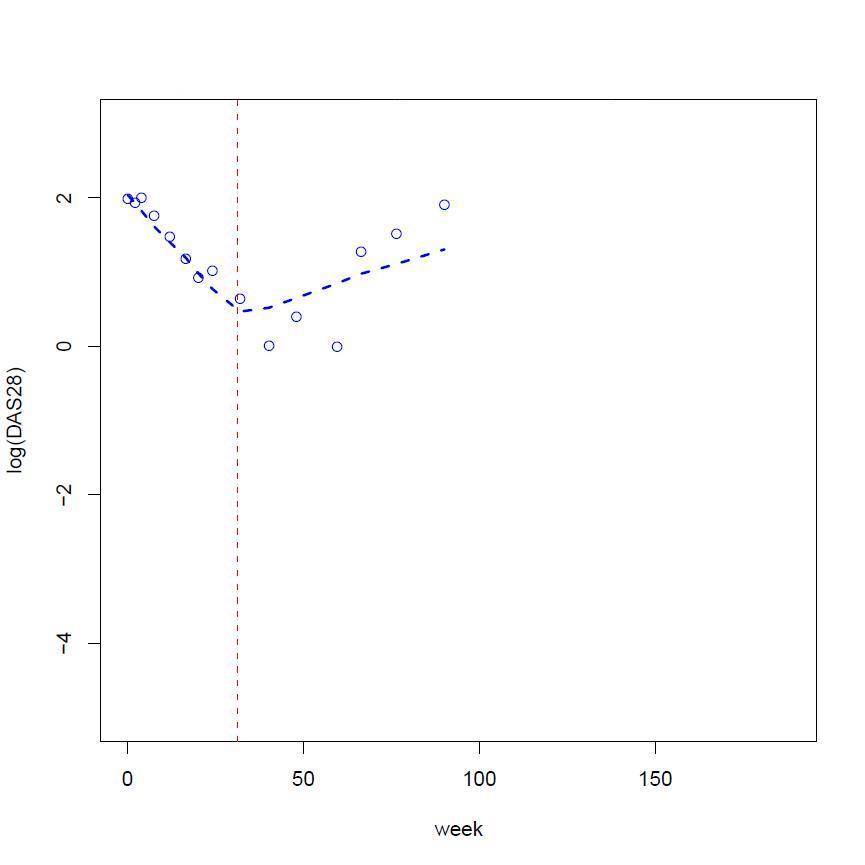}&\includegraphics[scale=.35]{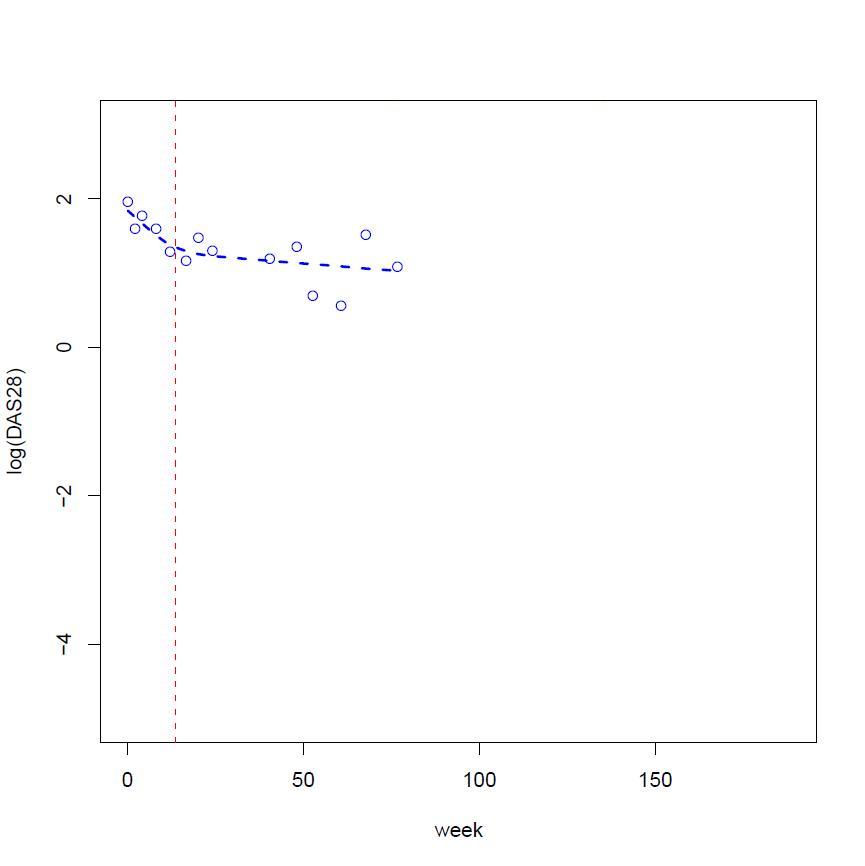}\\
\multicolumn{1}{l}{(c) Patient dropped out due to}&(d) Patient dropped out due to\\
\multicolumn{1}{l}{\ \ \ \ \ inefficacious treatment}&\ \ \ \ \  administrative reasons\\

\end{tabular}
\end{center}
\caption{
\label{fig:somepatients}
The fitted Model 6 to four patient level data.
The y-axis represents DAS28 scores on the log scale and the x-axis is time in weeks. The vertical dashed lines represent the patient specific change-point.}
\end{figure}

\begin{figure}[h]
 \begin{center}
 \includegraphics[scale=.65]{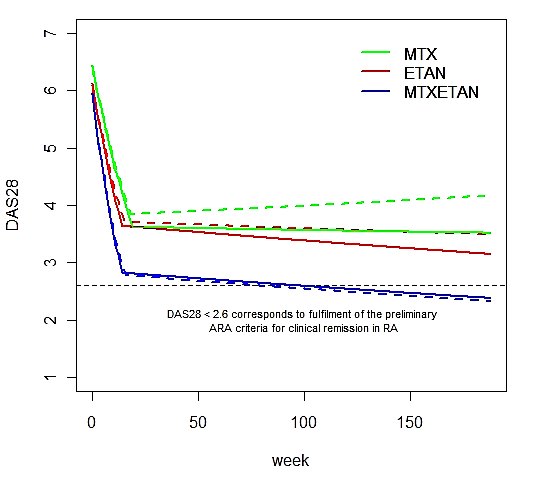}
 \end{center}
 \caption{\label{fig:model1vs6} Separate analysis Model 1 (solid lines) versus Joint analysis Model 6 (dashed lines)}
\end{figure}

Visually, Model 6 fits well to the subject-level data (Figure \ref{fig:somepatients}). Figure \ref{fig:model1vs6} 
shows the population-level curves for each
treatment group using separate analysis (Model 1) versus joint analysis (Model 6). One would conclude that subjects
exposed to the combo-therapy are more likely to reach clinical remission earlier (DAS28 score $< 2.6$). However, Model
1 tends to overestimate treatment effectiveness in the methotrexate group (MTX) where dropout due to inefficacy was the
highest 18.4\% compared to 17.5\% in the etanercept group (ETAN) and 5.6\% in the combo-therapy (MTXETAN).


\FloatBarrier
\section{Discussion}
Clinical trials are prone to incomplete outcome data due to patient drop. In the presence of informative dropout and when
drop out is differential across the treatment groups, separate analysis of the longitudinal data may lead to biased results. In
the example TEMPO study, separate analysis tends to overestimate the treatment effect in the methotrexate group (MTX)
where dropout due to inefficacy was the highest 18.4\% compared to 17.5\% in the etanercept group (ETAN) and 5.6\% in the combo-therapy (MTXETAN). The proposed joint model provides a way to perform longitudinal data analysis while
adjusting for informative dropout with competing risk. Such approach of joint modeling can make the effect estimates
less biased and therefore we can draw more reliable conclusions from the clinical trial. In the case where the proportion
of dropout is small and there is no association between dropout and patient outcome, the joint model analysis should give
similar results to the separate model analysis.

In this paper, we also considered a linear regression random-change point model for the longitudinal DAS28 scores to
account for the observed change-point pattern in the DAS28 scores with a steep decrease in DAS28 scores after initial
treatment followed by stability or a more gradual change. The complexity of the model is easily handled under a Bayesian
framework. The time to run the model is within minutes and convergence was observed for all parameters. Our proposed
Bayesian joint model can also be extended to safety endpoints where safety endpoints are modeled as a survival process.

\bibliographystyle{chicago}

\bibliography{longitudinalDAS28}






\subsection*{Appendix}
WinBUGS code for Model 6: Joint model that shares parameters from the longitudinal trajectory function
\begin{small}
\begin{verbatim}
########################################################################
# Data list requires the following:
# N = number of subjects in the study
# n[i] =  number of das28 scores observed for subject i
# Y[pointer[i]+j] = vector of log das28 scores. Uses a pointer to identify
#                  subject i and observation j
# pointer[i] = vector that points to the first observation of subject i in
#             the Y vector
# week[pointer[i]+j] = vector of time in weeks for for subject i and das28
#                     score observation j
# x1[i] = dummy variable that equals 1 if subject i received methotrexate
#        alone, 0 otherwise
# x2[i] = dummy variable that equals 1 if subject i received etanercept
#        alone, 0 otherwise
# dropoutweek.EFFY[i] = time of dropout due to inefficacious treatment in
#                      weeks for subject i. Use 'NA' if subject i is right
#                      censored
# censored.EFFY[i] = time of right censored in weeks for subject i for
#                    dropout due to inefficacious treatment. Use '0' if
#                    dropout due to inefficacious treatment was observed
#                    for subject i
# dropoutweek.AE[i] = time of dropout due to adverse events in weeks for
#                     subject i. Use 'NA' if subject i is right censored
# censored.AE[i] = time of right censored in weeks for subject i for dropout
#                  due to adverse event. Use '0' if dropout due to adverse
#                  event was observed for subject i
########################################################################

model{
  for(i in 1:N){
     for(j in 1:n[i]){
     ######################################################################
     # Longitudinal process with two lines that meet at a random 
     #change point
     ######################################################################
        Y[pointer[i]+j]~dnorm(y.star[pointer[i]+j],tau.y)
        y.star[pointer[i]+j]<-alpha[i]+beta[i,J[pointer[i]+j]]*
                              (week[pointer[i]+j]-change.point[i])
        J[pointer[i]+j]<-1+step(week[pointer[i]+j]-change.point[i])
     }
     # reparameterization so that alpha.baseline represents baseline score
     alpha[i]<-alpha.baseline[i]+beta[i,1]*(change.point[i])

     ######################################################################
     # Competing risk for informative dropout where dropout times are
     # bounded by censoring times and linked to the longitudinal process by
     # sharing parameters from the longitudinal trajectory
     ######################################################################
     dropoutweek.EFFY[i]~dnorm(theta.EFFY[i],
                         tau.EFFY[trtgrp[i]])I(censored.EFFY[i],)
     theta.EFFY[i]<-phi.EFFY[1]+phi.EFFY[2]*x1[i]+phi.EFFY[3]*x2[i]+
                  omega.EFFY[1]*alpha.baseline[i]+omega.EFFY[2]*beta[i,1]+
                  omega.EFFY[3]*beta[i,2]
     dropoutweek.AE[i]~dnorm(theta.AE[i],
                               tau.AE[trtgrp[i]])I(censored.AE[i],)
     theta.AE[i]<-phi.AE[1]+phi.AE[2]*x1[i]+phi.AE[3]*x2[i]+
                  omega.AE[1]*alpha.baseline[i]+omega.AE[2]*beta[i,1]+
                  omega.AE[3]*beta[i,2]
     trtgrp[i]<-1+step(x1[i]-0.5)+2*step(x2[i]-0.5)
  }
  ########################################################################
  # Priors for the parameters in the longitudinal and the dropout process
  ########################################################################
  for(i in 1:N){
     change.point[i]~dnorm(mu.changepoint[i], tau.changepoint)
     mu.changepoint[i]<-gamma.chgpt[1]+gamma.chgpt[2]*x1[i]+
     gamma.chgpt[3]*x2[i]
  }

  for(i in 1:N){
     alpha.baseline[i]~dnorm(mu.alpha[i],tau.alpha)
     mu.alpha[i]<-gamma.alpha[1]+gamma.alpha[2]*x1[i]+
                         gamma.alpha[3]*x2[i]  
  }

  for(i in 1:N){
     for(m in 1:2){
        # beta[i,1] and beta2[i,2] represent slopes 
        #before and after change point
        beta[i,m]~dnorm(mu.beta[i,m],tau.beta[m])
        mu.beta[i,m]<-gamma.beta[1,m]+gamma.beta[2,m]*x1[i]+
                             gamma.beta[3,m]*x2[i]
     }
  }
  ########################################################################
  # Hyper-priors (treatment effect parameters)
  ########################################################################
  for(trt in 1:3){
     gamma.alpha[trt]~dnorm(0,0.001)
     for(m in 1:2){
        gamma.beta[trt,m]~dnorm(0,0.001)
     }
     gamma.chgpt[trt]~dnorm(12, 0.01)
     phi.EFFY[trt]~dnorm(0,0.001)
     phi.AE[trt]~dnorm(0,0.001)
  }
  #######################################################################
  # Component for the joint analysis
  # omega.k[1]= coefficient of association between dropout and 
  #             baseline das28 scores
  # omega.k[2] = coefficient of association between dropout and 
  #              change in das28 scores before change point
  # omega.k[3] = coefficient of association between dropout and
  #              change in das28 scores after change point
  #######################################################################

  for(v in 1:3){
     omega.EFFY[v]~dnorm(0,0.001)
     omega.AE[v]~dnorm(0,0.001)
  }

  #######################################################################
  # Variance (precision) parameters
  ## ####################################################################
  tau.changepoint~dgamma(0.01, 0.01)
  tau.y~dgamma(0.01,0.01)
  tau.alpha~dgamma(0.01,0.01)
  tau.beta[1]~dgamma(0.01,0.01)
  tau.beta[2]~dgamma(0.01,0.01)
  for(grp in 1:3){
    tau.EFFY[grp]~dgamma(0.01,0.01)
    tau.AE[grp]~dgamma(0.01,0.01)
  }
}
\end{verbatim}
\end{small}

\end{document}